\documentclass{emulateapj}

\usepackage{graphicx}
\usepackage{natbib}
\usepackage{multirow}
\shorttitle{Probing the Existence of a Dark Matter Isothermal Core}
\shortauthors{Il\'\i dio Lopes \&  Joseph Silk}

\begin{document}

\title{Probing the Existence of a Dark Matter \\ Isothermal Core Using Gravity Modes}

\author{Il\'\i dio Lopes\altaffilmark{1,2,5}, Joseph Silk\altaffilmark{3,4,6}}

\altaffiltext{1}{Centro Multidisciplinar de Astrof\'{\i}sica, Instituto Superior T\'ecnico, Av. Rovisco Pais, 1049-001 Lisboa, Portugal} 
\altaffiltext{2}{Departamento de F\'\i sica, Universidade de \'Evora, Col\'egio Luis Ant\'onio Verney, 7002-554 \'Evora - Portugal} 
\altaffiltext{3}{Beecroft Institute for Particle Astrophysics and Cosmology, Department of Physics, University of Oxford, United Kingdom} 
\altaffiltext{4}{Institut d'Astrophysique de Paris, France} 
\altaffiltext{5}{E-mail: ilidio.lopes@ist.utl.pt} 
\altaffiltext{6}{E-mail: silk@astro.ox.ac.uk}  

\begin{abstract}
Although helioseismology has been used as an effective tool for studying the physical mechanisms acting in most of the solar interior,
the microscopic and dynamics of the deep core is still not well understood. Helioseismological anomalies may be partially resolved if the Sun captures light, non-annihilating dark matter particles, a currently discussed dark matter candidate that is motivated by recent direct detection limits. 
Once trapped, such particles (4-10 GeV) naturally fill the solar core. 
With the use of a well-defined stellar evolution code that takes into account an accurate description of the capture of dark matter
 particles by the Sun, we investigate the impact of such particles in its inner core.
Even a relatively small amount of dark matter particles in the solar core will leave an imprint  on the absolute frequency values of gravity modes,  
as well as the equidistant spacing between modes of the same degree. 
The period separation for gravity modes could reveal changes of up to 3\% for annihilating dark matter and of up to
20\% for non-annihilating  dark matter.  This effect is most pronounced in the case of the gravity dipole ($l=1$) modes.
\end{abstract}
\keywords{elementary particles --- dark matter --- Sun: interior  --- Sun: helioseismology  --- stars: interiors}

\section{Introduction}

Evidence for the existence of dark matter in the Universe is well established by cosmological observations 
\citep{art-Komatsuetal2009},  and its influence is also necessary in order to explain the formation of the structure 
of the current epoch Universe \citep{art-Springel2005Natur.435..629S}.
Various studies suggest that dark matter is constituted of massive, 
neutral, weakly interacting non-baryonic particles. Furthermore, independent considerations from particle physics also
propose the existence of such particles \citep{rev-BertoneHS2005}. 

Recently, the results from  several particle physics detection experiments
\citep{art-Bernabeietal2010AIPC.1223...50B,art-CDMSII2010science,art-Aalseth2010,art-Fitzpatrick2010}
have been interpreted in terms of  weakly interacting massive dark matter particles (WIMPs)   with a mass smaller 
than 10 GeV, and with spin-dependent elastic scattering cross-sections with baryons as 
large as  $10^{-32} {\rm cm^2}$  \citep{art-XENON10_SD2008,art-COUPP2008})  
or spin-independent elastic scattering cross-section with baryons of the order of $10^{-40} {\rm cm}^2$ 
\citep{art-CDMSII_SI2009}.
 The relatively weak experimental limits on spin-dependent interactions of dark matter particles are especially of interest for the Sun 
 with its large abundance of protons, for which both  self-annihilating 
 cold dark matter particle scenarios \citep{art-Bergstrom2009NJPh} and non-annihilating  \citep{art-Frandsen2010PhRvL} have previously been studied.  

 Investigation of the effects of the accumulation of dark matter particles by the Sun may 
 therefore be an important complement to direct detection searches for light WIMPs.  
 These particles are trapped in the Sun's 
interior when they collide with nuclei and lose (linear) momentum, and drift into the Sun's core. 
Collisions of captured particles with the local baryons transfer and redistribute thermal energy, and lower the central temperature by a few percent. 
In thermal equilibrium, the kinetic energy of dark matter particles is balanced by the local gravitational potential 
\citep{art-SpergelPress1985}. An estimation of the radius of the dark matter core is given by
$r_x\sim \left(9 k T_c/4\pi G\rho_c m_p\right)\;\sqrt{m_p/m_x}$ where $m_p$ and $m_x$ are, respectively, the mass of the proton and the mass of the dark matter particle,  $T_c$ and $\rho_c$ are the central temperature and the central  density of the Sun's core, and  $k$ and $G$ are, respectively, the Boltzmann and  Newton gravitational constants. This expression approximately gives the radius of the dark matter core in the Sun's interior.  It follows that the more massive a dark matter particle, the smaller is the radius of the dark matter core, and the less important is the impact of dark matter in the evolution of the Sun.

The Sun shows a complex pattern of surface oscillations whose restoring forces are produced either 
by compressibility or buoyancy. The pressure perturbations give rise to acoustic sound
waves in the high-frequency part of the spectrum, and buoyancy variations drive gravity waves in the low-frequency range of the spectrum.  
The small amplitude surface perturbations observed in the Sun can be described as a sum of eigenstates. Each eigenstate has a spatial
counterpart that is defined by a spatial eigenfunction that depends on the thermodynamical structure of the background state 
(the Sun's internal structure),  and a time-dependent  eigenfunction that is characterized by the frequency $\nu_{n,l}$. 
The numbers $l$ and $n$ are positive integers, known as the degree and radial order of the mode\citep[e.g.,][]{art-Gough1993}.

During the last 50 years, accurate measurements of frequency values were obtained for more than 5 thousand 
acoustic mode frequencies $\nu_{l,n}$. This achievement is the result of the combined efforts of 
several observational networks, such as BISON\citep{art-BISON1996} and GONG \citep{art-GONG1996Sci},
followed by the helioseismic experiments from the SoHO mission: 
GOLF\citep{art-GOLF1995}, VIRGO\citep{art-VIRGO1995} and MDI-SOI \citep{art-MDI1995}.
Seismology has provided the most powerful tool to probe the Sun's interior to date. Accurate frequency measurements 
of large numbers acoustic modes have been made, 
including the radial and dipole global acoustic modes which  penetrate more deeply into the core. 
 Unfortunately, the diagnostics of the Sun's core provided by these  acoustic modes are still insufficient.  However,
 the possible discovery of gravity modes by current ongoing experiments\citep{art-Garcia2007Sci} 
 or future ones could be a breakthrough  towards fully understanding the physics of the Sun's core
 and establishing the possible existence of dark matter. 
 The impact of dark matter particles in the Sun's core \citep{art-Coxetal1990ApJ...353..698C,art-Dearbornetal1990,art-Kaplanetal1991}
 and its consequences on the acoustic spectrum 
 were previously analysed by \citet{art-LopesSH2002} among others,
 and more recently by 
\citet{art-Cumberbatch2010} and \citet{art-Taoso2010}.

In this Letter, we have computed a series of solar models evolving within
dark matter halos for which we have estimated the  impact of dark matter particles in the 
present-day Sun.
As a reference, we use an updated solar standard model that provides a seismic diagnostic similar 
to other solar standard models found in the literature. Furthermore, we have computed the gravity mode oscillation 
spectrum for such dark matter scenarios.
\begin{figure} 
\centering
\includegraphics[scale=0.5]{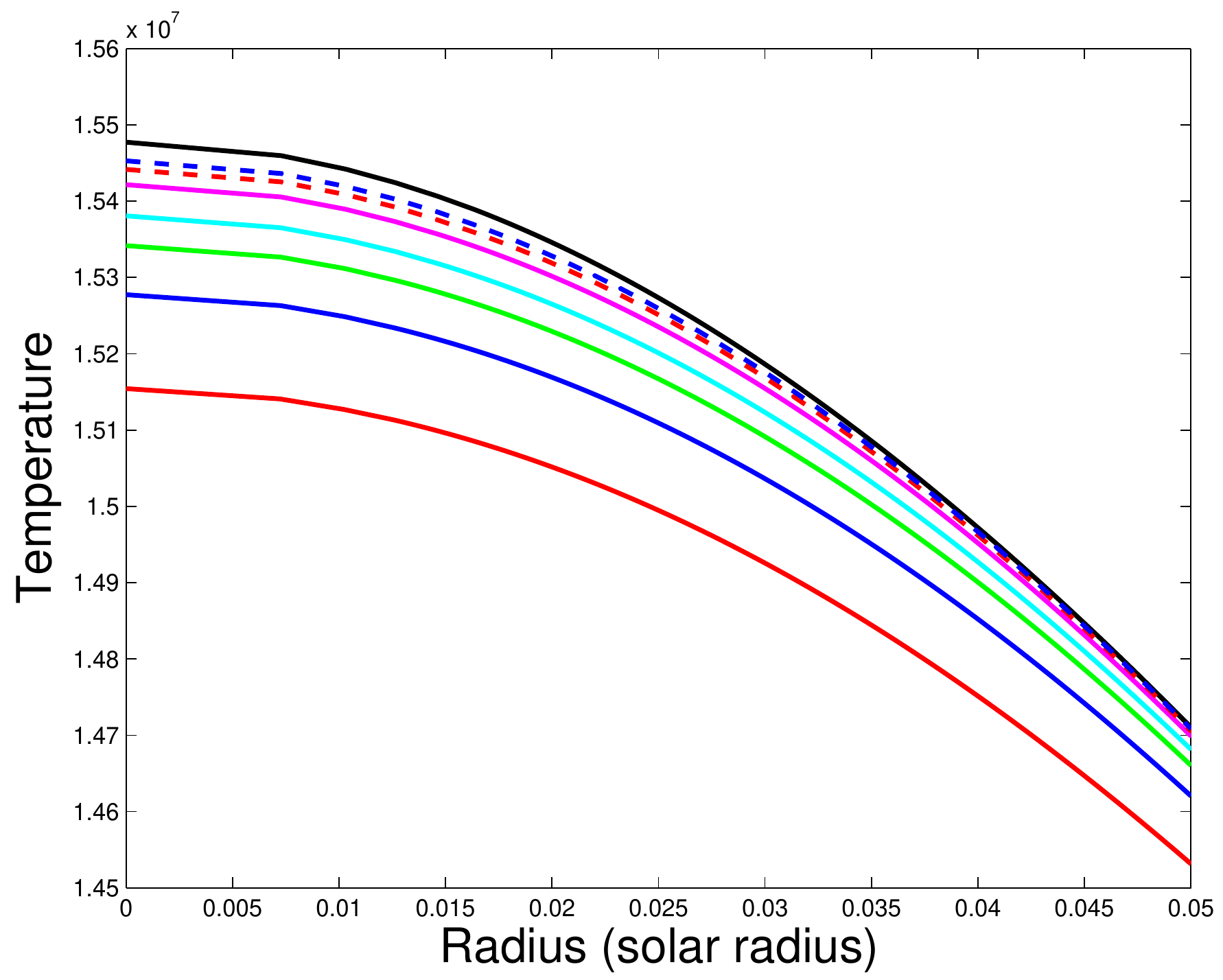}
\includegraphics[scale=0.5]{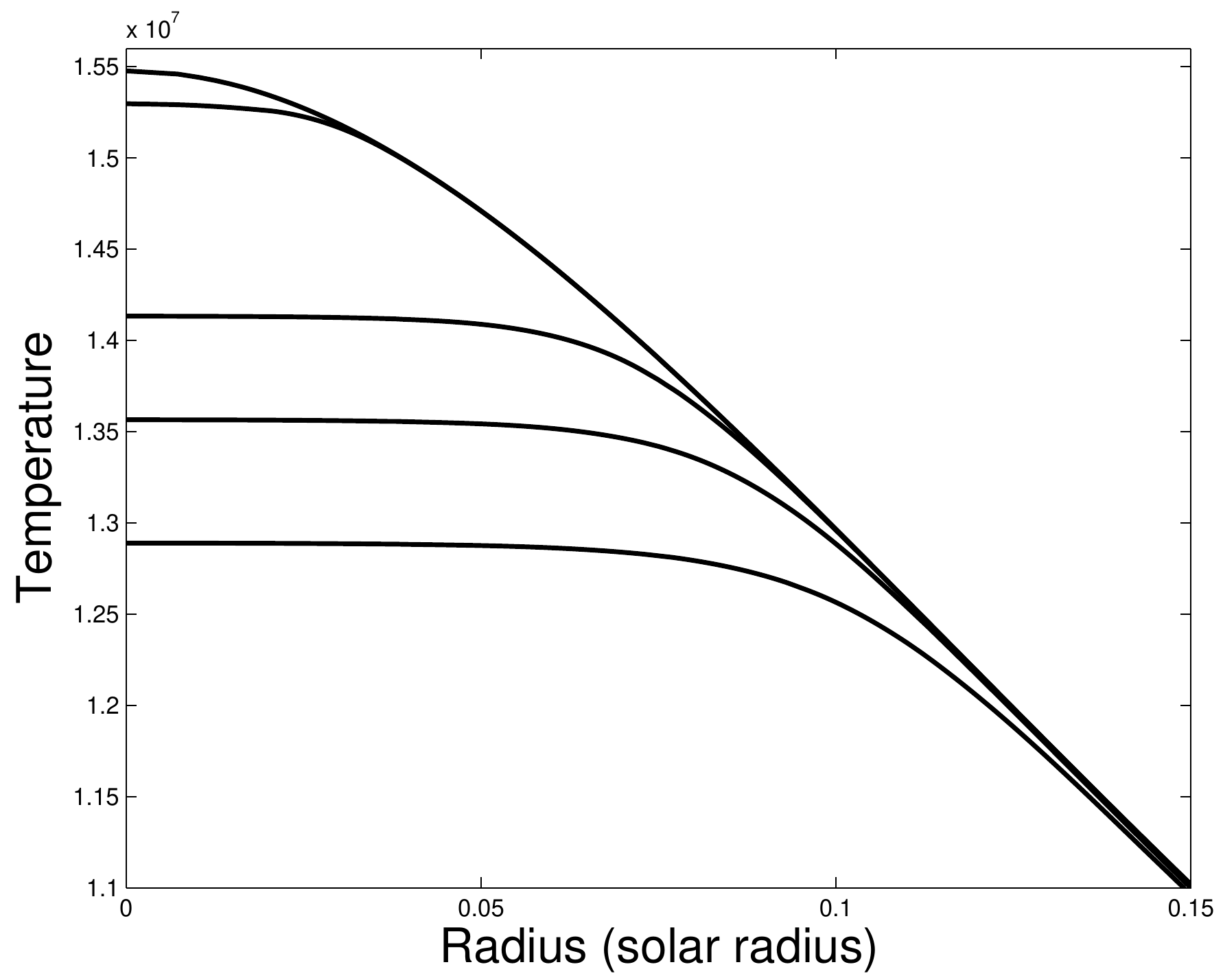} 
\caption{ 
Comparison of temperature profiles between the solar standard model and 
models of the Sun evolving in different dark matter halos.
 The halos are constituted of self-annihilating and non-annihilating massive particles with the following masses: 
 (i) self-annihilating particles (coloured curves): 3 Gev (red), 4 GeV (blue), 5 GeV (green), 6 GeV (cyan), 8 GeV (magenta), 
 10 GeV (dashed-red) and 12 GeV (dashed-blue);
 (ii) non- annihilating particles (black curves, from low to higher temperature):  5 GeV, 7 GeV, 10 GeV and 50 GeV.
The particles interact with baryons with a spin-dependent scattering cross-section of the order of
 $3\; 10^{-33} {\rm cm^2}$ (i: coloured curves) or $2\;10^{-35} {\rm cm^2}$ (ii: black curves), and spin-independent scattering 
 cross-sections of the order of $10^{-40} {\rm cm^2}$ (negligible). The product of the self-annihilation cross-section 
 and the relative velocity of colliding particles is of order of
 $3\;10^{-33} {\rm cm^3 s^{-1}}$ (i:coloured curves) or 
 $10^{-50} {\rm cm^3 s^{-1}}$ (ii:black curves;suppressed annihilation). 
 The black curve with the highest central temperature corresponds to the solar standard model.
 }
\end{figure}

\section{Helioseismology and the Sun's evolution in dark matter halos}

The capture rate of dark matter particles in the Sun was computed in detail by taking into account 
their scattering with the different chemical species of the solar plasma \citep{art-Gould1987,art-GondoloEdsjoDarkSusy2004,art-CL2010procIU,art-astrophCL}.
The presence of dark matter inside the star can change its evolution by two mechanisms: (i) changing the transport of
energy  inside the star \citep{art-SpergelPress1985,art-GouldRaffelt1990a},
or (ii) by contributing to the energy production that sustains the gravitational collapse of the star
 \citep{art-Ioccoetal2008,art-FreeseBSG08,art-CasanellasLopes2009}.
In the case of the Sun, the latter mechanism is negligible \citep{art-LopesSH2002}.
The efficiency of the energy transport by dark matter particles depends, among other factors, 
on the scattering cross-sections of dark matter particles on baryons. 

Dark matter particles might have a total scattering cross-section  $\sigma_t$ with several components: spin-independent
interactions with all chemical elements in the Sun, and spin-dependent interactions with hydrogen.  The presence of such particles
contributes to the local energy transport.  The efficiency of the heat transport by dark matter particles is dependent on the value
$\sigma_t$, relative to the fiducially critical cross-section $\sigma_c = (m_p/M) R^2\sim$ $8\;10^{-36} {\rm cm^2}$, 
where $m_p$ is the proton mass and $M$ and $R$ are the mass and radius of the Sun. The optimal heat transport depends 
on the dark matter particle  scattering cross-section $\sigma_t$. For particles with small mean free paths (diffusion regime)
or with large cross-sections ($\sigma_t > \sigma_c$), the conductivity falls as $\sigma_c/\sigma_t$. Similarly, for particles with large mean free path
(Knudsen regime) or lower cross-sections ($\sigma_t < \sigma_c$), the conductivity falls as $\sigma_t /\sigma_c$. 
Furthermore, a suppression factor is required in this latter transport regime \citep{art-GouldRaffelt1990a,art-GouldRaffelt1990b}.
Both regimes are included in our stellar evolution code  \citep{art-LopesSH2002,art-CasanellasLopes2009}.
The numerical computation of such effects on the transport of energy can be accomplished by means 
of changing the production of energy or by changing the coefficient for radiative transfer.
We have implemented both methods. The results obtained in both cases are similar. In this
work, the transport of energy by dark matter is computed as a coefficient for the radiative transfer.
This seems to be more realistic and is numerically more stable. 
Several models of evolution of the Sun were computed in different dark matter scenarios
of suppressed and non-suppressed annihilation dark matter particles.
 Our reference model is an updated solar standard model  \citep{art-TL1993,art-AsplundGrevesseSauval2005}
that shows an acoustic seismic diagnostic similar to other solar standard models 
\citep{art-TC2004PhRvL,art-BSB2005,art-Guzik2010,art-TC2010}. 
We include self-annihilating and non-annihilating massive particles with the following masses: 
self-annihilating particles from  3-12 GeV  and 
non- annihilating particles from  5-50 GeV.
The particles interact with baryons with a spin-dependent scattering cross-section 
 $3\; 10^{-33} {\rm cm^2}$ or $2\;10^{-35} {\rm cm^2}$, and  a spin-independent scattering 
 cross-sections of $10^{-40} {\rm cm^2}$ (in fact, this is negligible). The product of the self-annihilation cross-section 
 and the relative velocity of colliding particles taken to be 
 $3\;10^{-33} {\rm cm^3 s^{-1}}$  or 
 $10^{-50} {\rm cm^3 s^{-1}}$ (for the case of suppressed annihilation).

\begin{figure}[!thbp]
\centering
\includegraphics[scale=0.38]{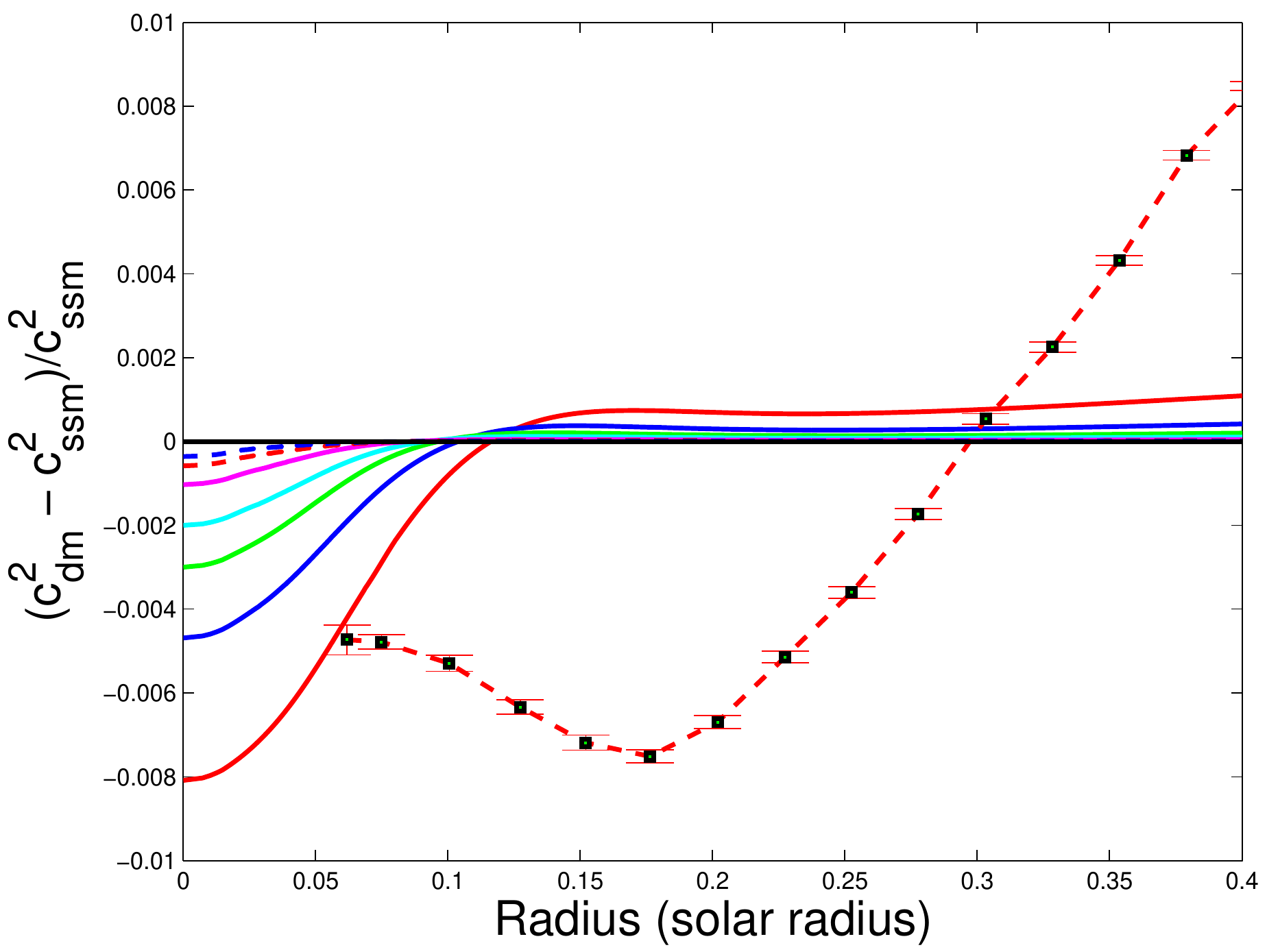}
\includegraphics[scale=0.38]{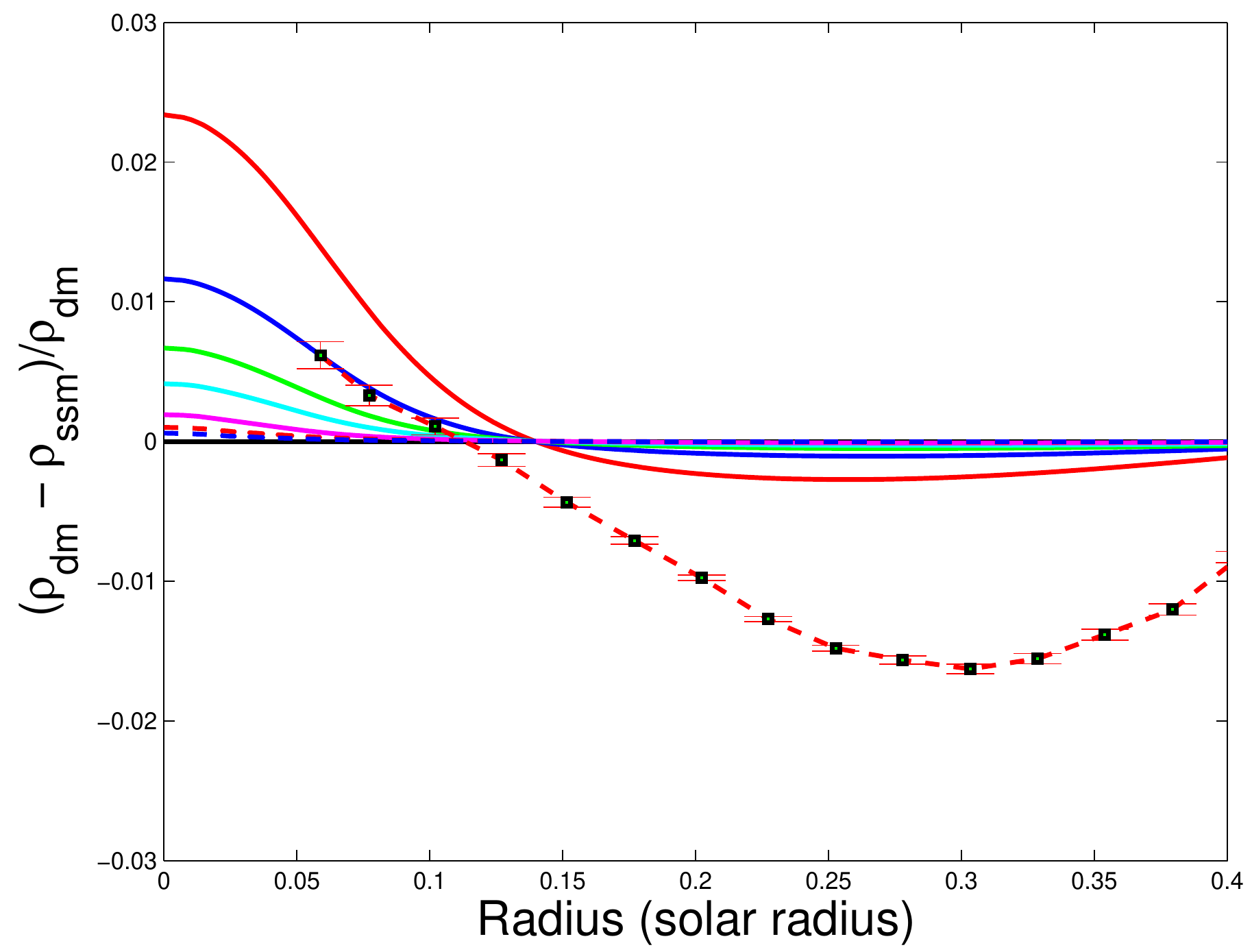}
\caption{Comparison between the solar standard model and different solar models evolved 
within self-annihilating dark matter halos: (a) sound speed radial profile and (b) density radial profile. 
The red dotted curve corresponds to the difference between inverted profiles and our solar
 standard model (see text). See caption of Figure 1 for the details on self-annihilating scenarios (case i: coloured curves).}
\end{figure}

In the case of non-annihilating particles, the numerical model shows a 2\% reduction of the 
central temperature of the Sun's core due to the lightest massive particles (Figure 1). 
This temperature reduction seems to be stronger than that obtained by \citet{art-Taoso2010}. 
Additionally, the square of the sound speed  and density profiles incur a variation of between 
0.8\% and 2\% (Figure 2). In the case of non-annihilating particles, the decrease in the central temperature 
is of the order of 8 \% for a particle with a mass of 12 GeV, and even larger for lighter particles. 

Figure 2 shows the profiles of sound speed and density in the case of annihilating dark matter scenarios 
compared with the inverted profiles. 
The inversion of the sound speed and density profiles were done using the seismic data of the BISON and GONG
networks \citep{art-Basu2009}. This seismic data is consistent with the previous high accuracy measurements 
done by the GOLF and MDI instruments of the SoHO mission \citep{art-TC2001ApJ...555L..69T}.
 It is evident that the present acoustic seismology is not able to
 probe the inner core of the Sun accurately.  In summary, our knowledge about the Sun's interior is quite accurate up to 10\% 
 of the Sun's radius\citep{art-TurckChieze2004,art-Garcia2007Sci}.  The dark matter is expected to accumulate in the inner core of the Sun.
 An accurate description of such deep layers of the Sun's interior can be obtained if gravity modes are observed.
 
\section{Gravity modes and the existence of a dark matter core}

\begin{figure} 
\centering
\includegraphics[scale=0.4]{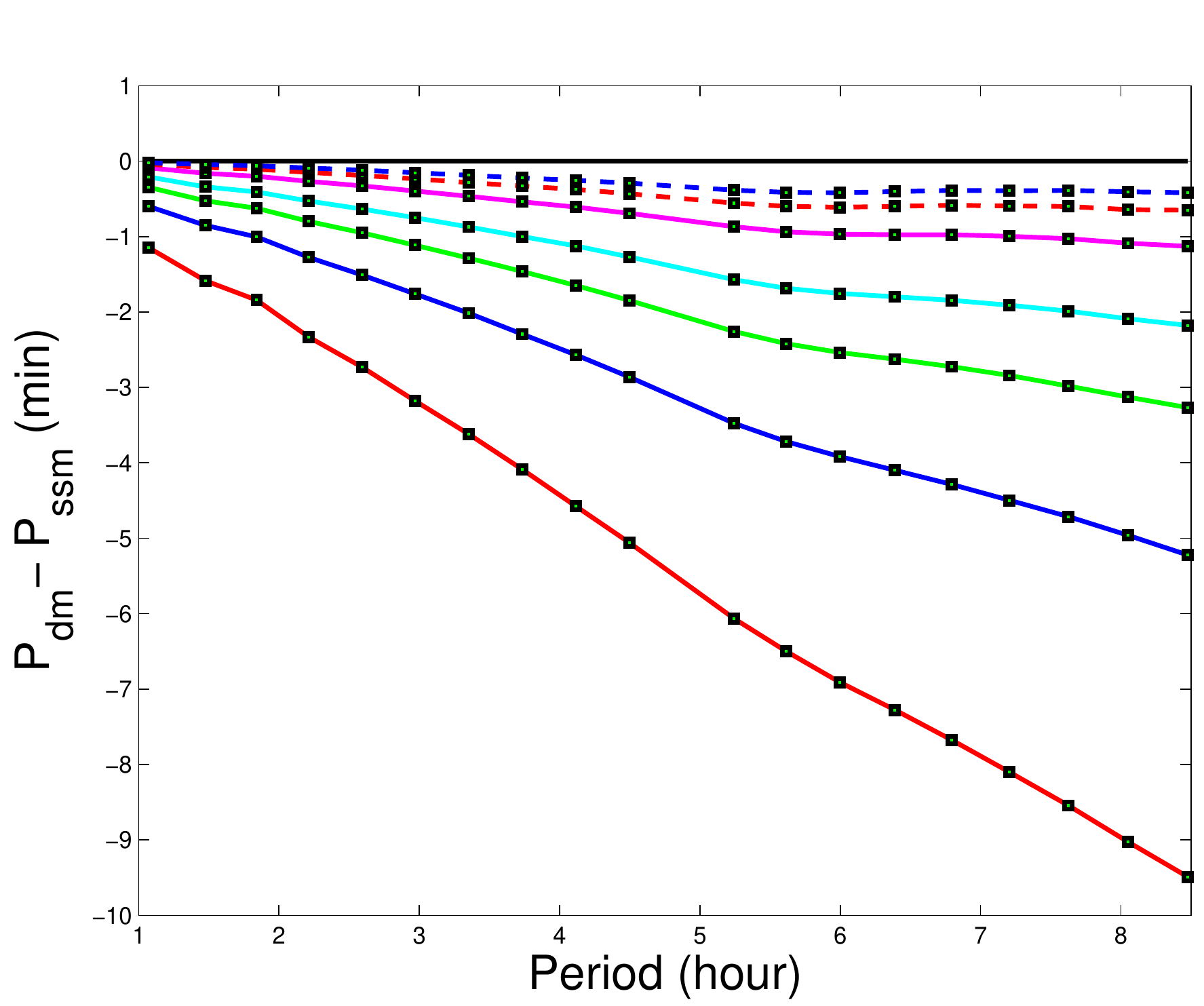}
\caption{Comparison  period  table $P_{l,n}$ between the solar standard model and models
 of the Sun evolving in different dark matter halos (scenario i). The periods $P_{l,n} $
 shown corresponds to the dipole gravity modes ($l=1$). This corresponds to dark matter halos constituted 
 of annihilating massive particles. (case i: coloured curves).  See caption of Figure 1 for the details 
 of the  self-annihilating scenario.  
}
\end{figure}

An additional probe of the Sun's interior, capable of finding small dark matter effects occurring in the Sun's core,
can be performed through gravity modes. 
We have computed the gravity modes of oscillation of the present Sun  for different dark matter 
halos scenarios \citep{art-Ch-Dals2008Ap&SS}. The changes
 in the spectrum of gravity modes, due to the existence of dark matter in the Sun's core are visible even in the case of 
 very small amounts of dark matter accumulated in its core. In the case of self-annihilating dark matter scenarios, 
 the period for gravity dipole ($l=1$) mode changes are from a few tens of seconds up to 10 minutes (see Figure 3). 
 In non-annihilating  dark matter scenarios, the effect is much stronger. For very light particles, the gravity dipole modes 
 could have their periods reduced by 80 minutes. 
This existence of a dark matter core also affects the period spacing of gravity modes, a quantity identical to the large separation
for acoustic modes. In principle, this should provide  the first clear indication of the existence of a gravity oscillation pattern in the observed spectrum.   
A qualitative expression for the large period separation can be obtained for the case of gravity 
modes with low degree $l$ and high-order $n$  modes, where the period $P_{l,n}  (=\nu^{-1}_{l,n})$ is given by  
 \begin{equation}
 P_{l,n} =\frac{P_o}{\sqrt{l(l+1)} }\; \left(n+\frac{l}{2}+\phi \right) +{\cal O}\left(\frac{1}{P_{l,n}}\right),
 \end{equation}
 with
\begin{equation}
 P_{o} =2\pi^2 \left[\int_0^R\frac{|N|}{r} dr \right]^{-1},
 \end{equation}
 where $N$ is the buoyancy, and $\phi$ is a phase term sensitive to the layers below the base of the convective zone 
 \citep{art-Tassoul1980ApJS}.
 This expression, and in particular $P_o$, tells us that gravity mode frequencies are determined by the density stratification of the core,
  through the buoyancy.   The value of $P_o$ is of the order of 34.10 minutes in the case of our solar standard model.
Usually, the value of $P_o$ is computed from the large period separation, $\delta P_{l,n}= P_{l,n}- P_{l,n-1}$.
In the case of dipole modes ($l=1$),  it reads  $\delta P_{1,n}=P_o/\sqrt{2}$.
In the case when the dark matter core becomes isothermal, the value of $P_o$ is strongly affected.
The structure differences between the solar standard model (ssm) and different dark matter scenarios (dm)
 can be estimated by computing large separation period differences,  which is equivalent to measure period differences,
 $\Delta P_{l,n}= P^{dm}_{l,n}- P^{ssm}_{l,n}$ ($l$ and $n$ fix).
 It follows  that $\Delta P_{l,n}/ P^{ssm}_{l,n}\approx \Delta P_{o}/ P^{ssm}_{o}$. 
The structure differences produced by the presence of dark matter in the Sun's core leads to a significant change
in the period separation. Figure 4 shows the large period separation for several self-annihilating dark matter scenarios
in the case of gravity dipole modes  ($l=1$). The period spacing in such models can experiment changes of up to 3 \%.
In the case of non-annihilating scenarios the period spacing can be reduced by as much as 20 \%.

\section{Conclusion}

\begin{figure} 
\centering
\includegraphics[scale=0.4]{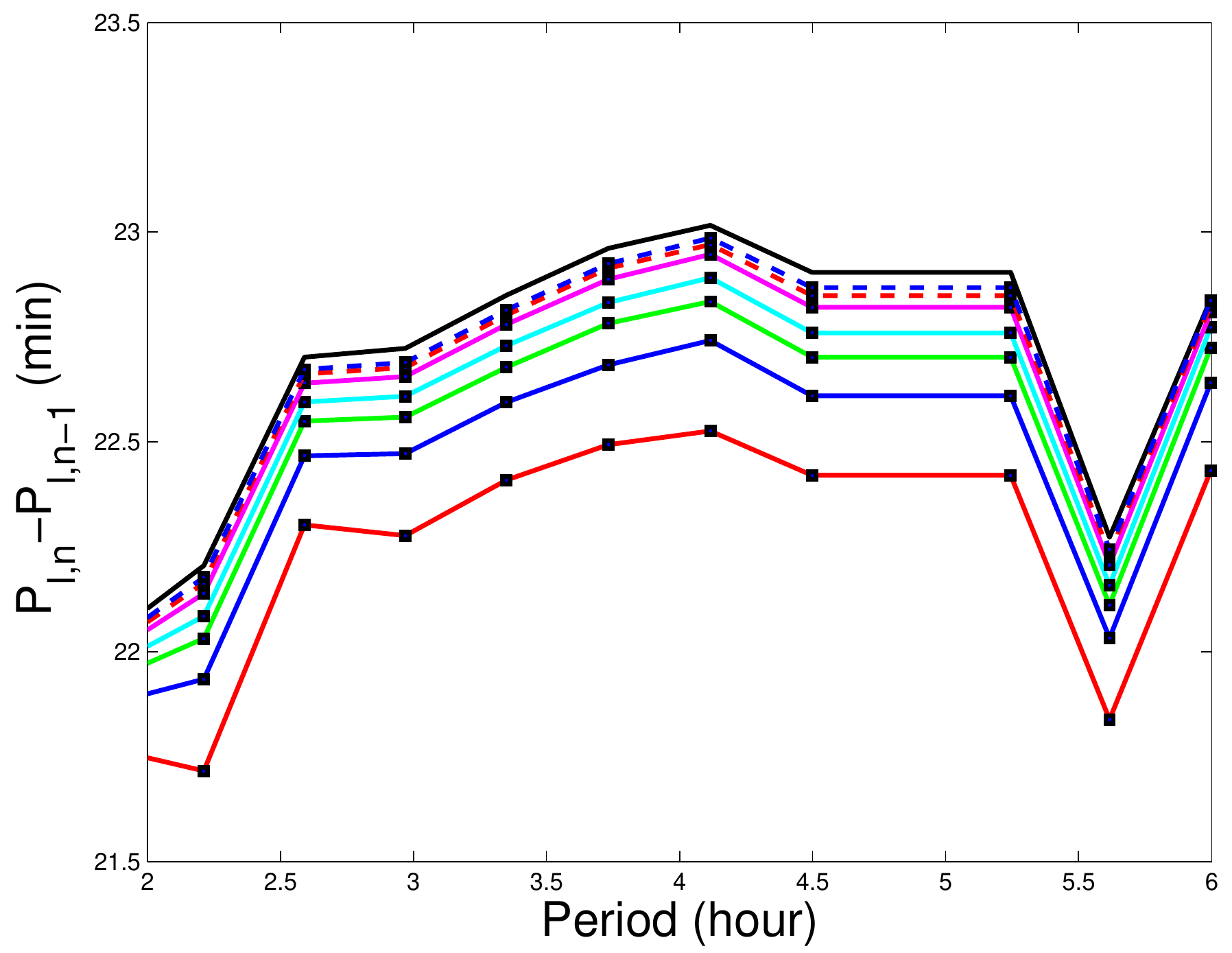}
\caption{The large separation for gravity modes: The large separation was computed from periods $P_{l,n}$, 
$\delta P_{l,n}=P_{l,n}-P_{l,n-1}\approx  P_{o}/ \sqrt{l(l+1)} $ shown in Figure 3. 
This was computed for dipole gravity modes ($l=1$). These dark matter halos are constituted of
 self-annihilating massive particles (case i: coloured curves). See caption of Figure 1 for details of the  self-annihilating scenario.}
\end{figure} 

The possible existence of a dark matter core, even in the case of non-annihilating dark matter scenarios of low mass particles, 
will be always contained  within the first 5-10\% of the solar radius. Such a dark matter core will very likely have a small temperature 
gradient where dark matter particles are present and, in the case of strongly interacting  dark matter scenarios, it will form a fully isothermal core. 
Such  an  isothermal core will produce two distinct signatures in the gravity modes: (i) by changing the frequency 
values of gravity modes, (ii) and by changing the large period separation between gravity modes of the same degree 
and consecutive radial order.

The impact of the dark matter in the evolution of the Sun and its impact in the helioseismology acoustic data as well as in the solar neutrinos observables
 has been addressed by several authors\citep[e.g.,][]{art-Dearbornetal1990,art-Kaplanetal1991,art-LopesSH2002,let-LopesSilk2002,art-LopesBS2002,let-LopesSilk2010}. 
 More recently a specific study has been done to explore a new class of fundamental particle candidates which are able to produce significant changes in the 
structure of core of the Sun \citep{art-Cumberbatch2010, art-Frandsen2010PhRvL,art-Taoso2010}.
In this study, we have explored in more detail how the gravity spectrum could be modified by the presence of such types of dark matter 
particles in the Sun's core.

Nevertheless, other physical processes participating in the evolution of stars need to be better understood in order to take full account 
of the effects caused by dark matter in the Sun's core. In particular, the new CNO composition has lead to a smaller central temperature 
than the one required by neutrino detection
 \citep[e.g.,][]{art-Guzik2010,art-TC2010}.  
The inclusion of well-known physical processes in the Sun's evolution, such as differential rotation, 
 meridional circulation, magnetic breaking, formation and evolution of the solar tachocline layer and solar dynamo\citep[e.g.,][]{art-Charbonneau2005,art-PassosLopes2008ApJ},  
 as well as the transport of angular momentum by gravity waves and/or magnetic fields (among other dynamical processes) will lead to minor 
 structure changes throughout the evolution of the star \cite[e.g.,][]{art-TC2010}. 
In some cases, such dynamical processes will increase the discrepancy between solar neutrino measurements and acoustic seismology. 
In other cases,  hydrodynamical processes will lead to reductions of the (small) differences between the solar model and observations \citep[e.g.,][]{art-Garaud2010ApJ}. 
Nevertheless, a full quantitative account of the physical processes involved in the evolution of the star is fundamental 
to finding the signature of dark matter in the Sun's core. 

The possible discovery of gravity modes by current space missions such as  SoHO or the new generation experiments SDO and PICARD 
will be of major importance in the search for dark matter inside the Sun. In addition to these, the GOLF-NG instrument  is specifically 
designed for a future solar mission. The goal of this instrument is to probe the very central region of the Sun, around 0.5\% 
of the solar radius, the most likely place to find dark matter.

We thank the anonymous referee for  advice in improving this paper.
We gratefully acknowledge the authors of CESAM (P. Morel), ADIPLS (J. Christensen-Dalsgaard), 
and DarkSusy (P. Gondolo, J. Edsj\"{o}, P. Ullio, L. Bergstr\"{o}m, M. Schelke and E. Baltz). 
This work was supported by grants from "Funda\c c\~ao para a Ci\^encia e Tecnologia" (SFRH/BD/44321/2008). 
\providecommand{\noopsort}[1]{}\providecommand{\singleletter}[1]{#1}%


\end{document}